\documentclass[reprint,prb,superscriptaddress]{revtex4-1}

\usepackage{graphicx}
\usepackage{amsmath, amsthm, amssymb}

\begin{document}

\title{Observation of Bloch oscillations in molecular rotation}

\author{Johannes Flo\ss}
\thanks{These authors contributed equally to this work.}
\affiliation{Department of Chemical Physics, Weizmann Institute of Science, 234 Herzl Street, Rehovot 76100, Israel}
\author{Andrei Kamalov}
\thanks{These authors contributed equally to this work.}
\affiliation{Stanford PULSE Institute, SLAC National Accelerator Laboratory, Menlo Park, California 94025, USA}
\author{Ilya Sh.~Averbukh}
\affiliation{Department of Chemical Physics, Weizmann Institute of Science, 234 Herzl Street, Rehovot 76100, Israel}
\author{Philip H.~Bucksbaum}
\affiliation{Stanford PULSE Institute, SLAC National Accelerator Laboratory, Menlo Park, California 94025, USA}
\date{\today}


\maketitle

\textbf{
The periodically kicked quantum rotor~\cite{casati06} is known for non-classical effects such as quantum localisation in angular momentum space~\cite{casati79,fishman82, kamalov15} or quantum resonances in rotational excitation~\cite{casati79,izrailev80}.
These phenomena have been studied in diverse systems mimicking the kicked rotor, such as cold atoms in optical lattices~\cite{raizen99,ibloch2005}, or coupled photonic structures~\cite{morandotti99,schwartz07,segev13}.
Recently, it was predicted~\cite{floss12,floss13,floss14} that several solid state quantum localisation phenomena -- Anderson localisation~\cite{anderson58}, Bloch oscillations~\cite{bloch29,zener34}, and Tamm-Shockley surface states~\cite{tamm32,shockley39}  -- may manifest themselves in the rotational dynamics of laser-kicked molecules.
Here, we report the first observation of rotational Bloch oscillations in a gas of nitrogen molecules kicked by a periodic train of femtosecond laser pulses.
A controllable detuning from the quantum resonance creates an effective accelerating potential in angular momentum space, inducing Bloch-like oscillations of the rotational excitation.
These oscillations are measured via the temporal modulation of the refractive index of the gas.
Our results introduce room-temperature laser-kicked molecules as a new laboratory for studies of localisation phenomena in quantum transport.
}

Bloch oscillations present one of the most famous and intriguing quantum localisation effects in solid state physics:
Electrons in crystalline solids subject to an external dc electric field exhibit oscillatory  motion instead of a mere uniform acceleration as in empty space.
This effect  was  predicted at the inception of quantum mechanics in 1929~\cite{bloch29,zener34}, but it took more than 60 years to first observe Bloch oscillations in semiconductor superlattice structures~\cite{feldmann92}.
Bloch oscillations were later observed in the momentum distribution of ultracold atoms driven by a constant force~\cite{dahan96}.
Recently, it was proposed to induce Bloch oscillations in the rotation of linear molecules excited by periodic trains of short nonresonant laser pulses~\cite{floss14}. This proposal is based on an analogy between coherent rotational excitation caused by the laser pulses and propagation of an electron in a one-dimensional periodic lattice.

A non-resonant, linearly polarised short laser pulse acts as a kick to a molecule.
It excites a rotational wave packet via multiple coherent Raman-type interactions~\cite{zon75,friedrich95b}:
$|\Psi\rangle(t)=\sum_J C_J(t) |J\rangle$, where $|J\rangle$ are the angular momentum states, and the projection quantum number $M$ is dropped since it is not changing.
 Under field-free conditions, the coefficients $C_J(t)$ oscillate as $\exp[-iBJ(J+1)t/\hbar]$, where $B$ is the molecular rotational constant.
The dynamics of the wave packet are determined by a single parameter, the rotational revival time $t_{\mathrm{rev}}=\pi\hbar/B$.
The wave packet revives exactly at integer multiples of the revival time~\cite{robinett04} when all the time-dependent phase factors become equal to unity.

Consider a train of laser kicks with a constant time delay $\tau$ between the pulses.
The time delay is chosen to be slightly detuned from the rotational revival time: $\tau=(1+\delta)t_{\mathrm{rev}}$.
Due to the detuning, each component of the rotational wave packet acquires a small phase $\phi_J=\pi\delta J(J+1)$ from pulse to pulse (integer multiples of $2\pi$ are dropped).
From now on, we follow the dynamics stroboscopically, by considering only $C_J(n)$, the wave packet coefficients just after the $n^{\text{th}}$ kick.
For weak pulses and small detuning, the change of $C_J(n)$ from the $n^{\text{th}}$ to the $(n+1)^{\text{th}}$ pulse is given as (see supplementary material)
\begin{multline}
C_J(n+1)-C_J(n)\approx\\i\frac{P}{4}\big[C_{J+2}(n)+C_{J-2}(n)\big]-i\phi(J)\ C_J(n) \,.
\label{eq.difference}
\end{multline}
The first term on the right hand side describes the laser coupling of the rotational levels, where $P=(\Delta\alpha/4\hbar)\int E^2(t)\mathrm{d}t$ is an effective strength of the laser pulse~\cite{averbukh01} ($\Delta\alpha$ is the molecular polarisability anisotropy, and $E(t)$ is the envelope of the laser electric field).
Note that the laser field only couples states of the same parity, $\Delta J=0,\pm2$.
The second term is caused by the detuning of the pulse train period from the rotational revival time $t_{\mathrm{rev}}$.
As the change of $C_J(n)$ is small, the difference equation~\eqref{eq.difference} can be recast as a differential equation, where $n$ becomes a continuous variable:
\begin{equation}
i\frac{\mathrm{d}C_J(n)}{\mathrm{d}n}=-\frac{P}{4}\big[C_{J+2}(n)+C_{J-2}(n)\big]+\phi(J)C_J(n) \,.
\label{eq.tdse}
\end{equation}
Equation~\eqref{eq.tdse} looks like the Schr\"odinger equation for a particle moving in a periodic 1D lattice, where the number of pulses $n$ plays the role of dimensionless time.
The sites of the lattice are the angular momentum states $|J\rangle$.
(Note that even and odd $J$ form two independent lattices.)
The first term on the right hand side of equation~\eqref{eq.tdse} describes the coherent hopping between lattice sites.
The second term can be interpreted as an on-site potential $V(J)=\phi(J)=\pi\delta J(J+1)$.

\begin{figure}
\includegraphics[width=\linewidth]{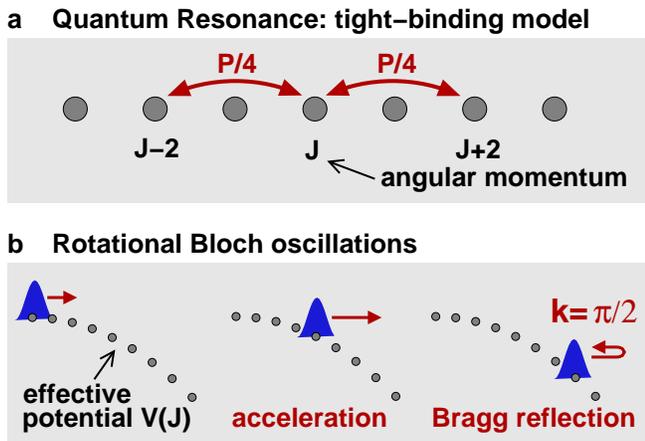}
\caption{
\label{fig.bo_sketch}
\textbf{The periodically kicked rotor as a particle in a 1D lattice.}
\textbf{a}, For a rotor kicked at the exact quantum resonance, the dynamics can be described by a tight-binding model.
The angular momentum levels $J$ form a discrete 1D grid, and the laser pulses couple sites with $\Delta J=\pm2$, where the coupling strength is proportional to the effective kick strength $P$ of the pulses (see text).
\textbf{b}, A detuning from the quantum resonance introduces an effective potential $V(J)$ to the model.
The dynamics are then similar to Bloch oscillations:
A wave packet (particle) is accelerated by the effective potential, and its wave vector $k$ grows.
When it reaches the edge of the Brillouin zone (here at $k=\pi/2$), the wave is reflected by Bragg reflection.
}
\end{figure}

At the quantum resonance, $\tau=t_{\mathrm{rev}}$ (i.e. $\delta=0$), the potential is $V(J)=0$.
In this case, equation~\eqref{eq.tdse} describes the standard tight-binding model in solid state physics~\cite{ashcroft76}, depicted in figure~\ref{fig.bo_sketch}~(a).
The eigenstates of this model are Bloch waves, which are characterised by the continuous quasimomentum $k$ and the energy dispersion relation $\varepsilon(k)$.
The latter is given as $\varepsilon(k)=-(P/2)\cos(2k)$.
Wave packets formed by these states propagate without limitation, reaching very large $J$ values.
This is the signature of the rotational quantum resonance effect~\cite{casati79,izrailev80}.

To treat the case of non-zero detuning, one can derive semi-classical equations of motion for the wave vector $k$ and the lattice coordinate $J$~\cite{ashcroft76}:
\begin{equation}
\frac{\mathrm{d}k}{\mathrm{d}n} =  -\frac{\mathrm{d} V(J)}{\mathrm{d} J} \,;\ \  \frac{\mathrm{d}J}{\mathrm{d}n} =\frac{\mathrm{d} \varepsilon (k)}{\mathrm{d} k}= P\sin(2k) \,.
\label{eq.diff}
\end{equation}
The first equation is Newton's second law, and the second one defines the group velocity of the Bloch waves in the lattice.
Equations~\eqref{eq.diff} can also be derived in a non-perturbative way for strong laser pulses using the formalism of $\epsilon$-classics~\cite{wimberger03}.
For negative detuning $\delta<0$, $V(J)$ is an accelerating potential [as depicted in figure~\ref{fig.bo_sketch}~(b)], and equations~\eqref{eq.diff} are similar to the ones describing electrons in crystalline solids subject to a constant electric field, the famous problem treated by Bloch and Zener~\cite{bloch29,zener34}.
For low $J$, the force $-\mathrm{d}V/\mathrm{d}J$ is weak, and the dynamics resembles the case of the quantum resonance.
As the Bloch wave packet moves to larger $J$, the quasimomentum $k$ grows and approaches the end of the Brillouin zone at $k=\pi/2$, where the length of the Bloch waves is comparable to the ``lattice constant'' $\Delta J=2$.
As a result, the wave is reflected due to Bragg scattering.
The now backwards moving wave packet is decelerated by the potential, until it stops and the cycle starts again.

\begin{figure}
\includegraphics[width=\linewidth]{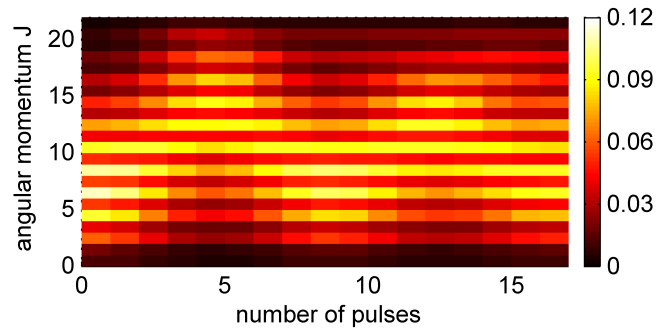}
\caption{
\label{fig.bo_simulation}
\textbf{Bloch oscillations in the angular momentum of laser kicked $^{14}$N$_2$ molecules.}
Simulated population of the angular momentum levels $J$, as a function of the number of laser pulses.
The initial rotational temperature is 298~K.
The considered pulse train parameters are $\tau=8.36~\text{ps}$ (0.2\% less than the revival time of $8.38~\text{ps}$) and an effective interaction strength of $P=5$ (see text).
}
\end{figure}

To illustrate the above predictions, we simulated the population distribution of the rotational levels $J$ for $^{14}$N$_2$ molecules at room temperature, interacting with a train of pulses with $P=5$.
The detuning is equal to $\delta=-0.2\%$.
The result of the simulation is shown in figure~\ref{fig.bo_simulation}.
One can clearly see the predicted oscillations of the rotational excitation.
For the first four pulses, the angular momentum population shifts to higher $J$; the shift per pulse is constant.
From the the fifth pulse on, the movement is reversed and directed towards lower $J$.
After eight pulses, the system returns approximately to the initial state, and the cycle starts again.

To observe the rotational Bloch oscillations experimentally, we employ a scheme similar to the one proposed earlier~\cite{floss14}.
A triple nested interferometer set-up is used to create a periodic train of eight femtosecond laser pulses~\cite{cryan09}.
These pulses interact with nitrogen molecules at room temperature and standard pressure conditions.
The rotational excitation of the molecules leads to molecular alignment, which causes optical birefringence.
We measure the time-dependent alignment signal and take its average over one rotational revival time as a measure of the rotational excitation (see Methods).

\begin{figure}
\includegraphics[width=\linewidth]{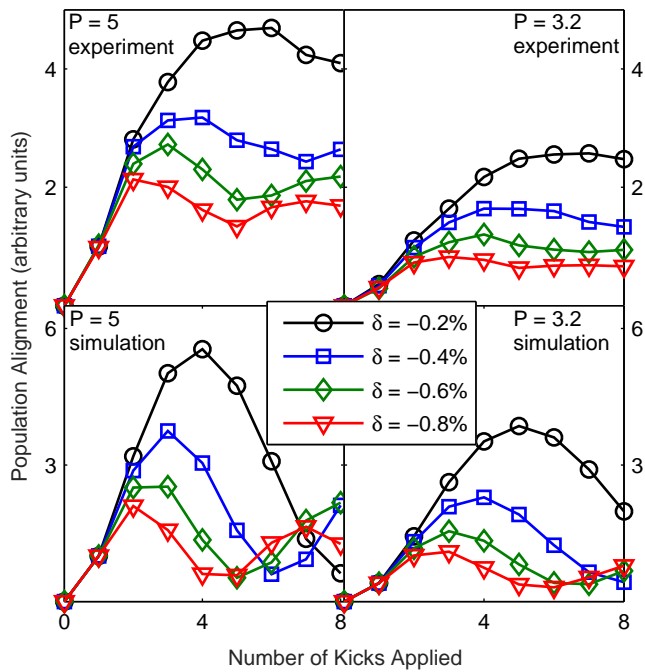}
\caption{
\label{fig.result}
\textbf{Measured and simulated time-averaged alignment signal.}
A comparison of the measured and simulated time-averaged alignment signals for periodically kicked nitrogen molecules, for different values of the kick strength $P$ and the detuning $\delta$.
The statistical error of the experimental data is much less than the plot symbols and therefore not plotted.
}
\end{figure}

Figure~\ref{fig.result} shows the results for several pulse train parameters $P$ and $\delta$, and demonstrates the predicted oscillatory behaviour.
The following trends can be observed:
First, the amplitude of the oscillations increases with the effective kick strength $P$, but decreases when the detuning $|\delta|$ is increased.
In addition, the oscillation period decreases both with increasing $P$ and $|\delta|$.
The same trends are found for the solutions of equations~\eqref{eq.diff}.
The statistical uncertainties are smaller than the plot symbols.

\begin{figure}
\includegraphics[width=\linewidth]{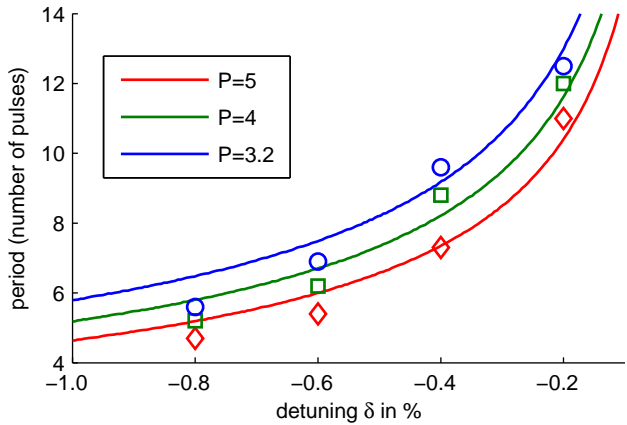}
\caption{
\label{fig.period}
\textbf{Period of the Bloch oscillation.}
The oscillation period as a function of the detuning, for different effective interaction strengths $P$.
The lines are the solution of the semi-classical model, equations~\eqref{eq.diff}, the markers are experimental values.
}
\end{figure}

To further compare the predictions with the experiments, we solved equations~\eqref{eq.diff}, choosing $J(0)=0$ and $k(0)=\pi/4$ as initial conditions.
This value of the quasimomentum corresponds to an initial growth rate of $dJ/dn=P$ of the angular momentum (the typical change of $J$ by a single pulse).
Figure~\ref{fig.period} shows the oscillation period as a function of the detuning $\delta$ for different kick strengths.
The curve predicted by the semi-classical model (solid lines) and the values extracted from the experiment (markers) show general agreement.
The visibility of the measured oscillations is reduced compared to the simulations presented in figure~\ref{fig.result}.
We attribute this to collisional decoherence as well as to imperfections of the pulse train.

Summarising, we described the first observation of Bloch oscillations in a quantum rotational system.
We used nitrogen molecules at room temperature and standard pressure conditions, kicked by a periodic train of femtosecond laser pulses, close to the quantum resonance condition.
Our results demonstrate that laser kicked molecules are subject to the same localisation phenomena that are seen in quantum transport.

\section*{Methods}

\begin{figure*}[t]
\includegraphics[width=\linewidth]{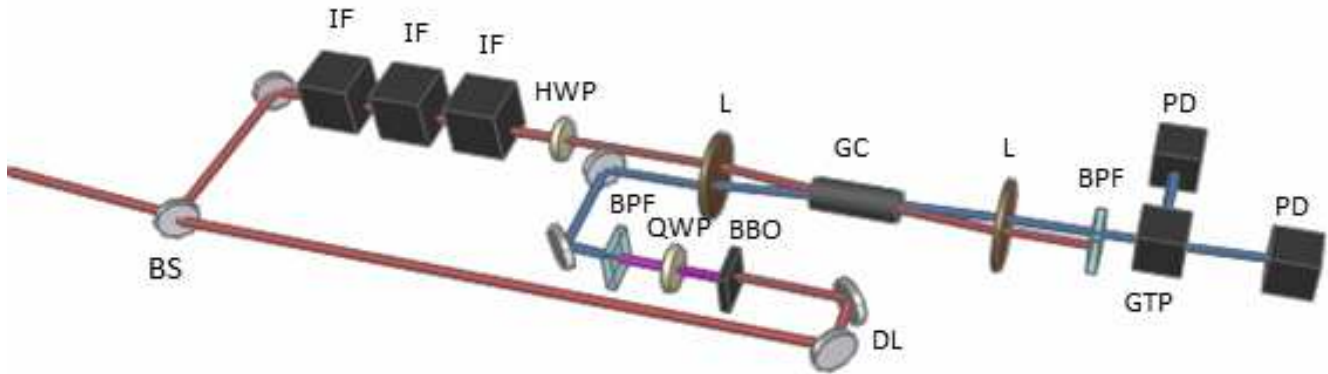}
\caption{
\label{fig.setup}
\textbf{Simplified sketch of the experimental set-up.}  A Ti:Sapph source generates an 800 nm pulse that enters a pump-probe set-up.  The pump is split into an eight pulse train using three nested interferometers.  The probe pulse is converted to circularly polarised 400 nm with variable path length.  The beams are focused using an off-axis parabola (depicted as a lens).  Molecular alignment triggered by the strong pump train causes birefringence that alter the circular polarisation of the weak probe pulse.  The probe's polarisation is split and measured, yielding a time-dependent molecular alignment signal that can be time-averaged to yield population alignment.  The abbreviations used are: BS - beam Splitter, IF - interferometer, HWP - half wave plate, L - lens, DL - automated delay line, BBO - $\beta$-BBO crystal, QWP - quarter wave plate, BPF - blue pass filter, GC - gas cell, GTP - Glan-Taylor polariser, PD - photodiode.
}
\end{figure*}

\textbf{Experiments.}
The data are measured using a standard ultrafast pump probe set-up, depicted in figure~\ref{fig.setup}.
A triple-nested interferometer set-up is utilised to split a single 3~mJ 70~fs 800~nm pulse into an periodic eight pulse pump train, as well as a weak probe pulse as described previously~\cite{cryan09}.
This train is focused by an off-axis parabola into a fused silica cell containing a constant flow of dry $^{14}$N$_{2}$ gas.
The induced rotational dynamics are encoded in the time-dependence of the angular alignment of the molecules (for recent reviews on laser molecular alignment see~\cite{ohshima10,fleischer12,pabst13}), and gives rise to a weak optical birefringence.
This time-dependent birefringence is probed using a circularly polarized 400 nm probe pulse with a variable optical path length.
The birefringence within the sample causes the probe's polarisation to become weakly elliptical after propagation.
The probe is split with a Glan-type polariser, and the polarisation component amplitudes are recorded as a function of probe delay time.
This measurement is performed with the pump train both present and blocked in rapid succession, and the normalised difference is proportional to the molecular alignment signal at the selected time delay between the pump and probe arms.
A detailed description of the set-up is presented in precedent work~\cite{cryan09}.

We use the fact that the time-averaged alignment is effectively a monotonic function of the angular momentum $J$, especially at high temperatures~\cite{kamalov15,leibscher04}.
An increase (decrease) in the expectation value $\langle J \rangle$ of the angular momentum translates into an increase (decrease) of the time-averaged alignment.
Therefore, one can observe the Bloch oscillations indirectly via the time-averaged alignment signal (population alignment).
Each value of the population alignment is found by averaging the molecular alignment signal over one rotational revival time; this signal is sampled at about 100 uniformly spaced probe delay times.
The probe times are arranged to accurately show the population alignment and avoid contamination by the coherent alignment, which has different decoherence mechanisms.
Each population alignment value is measured over $150,000$ laser shots and has negligible statistical error.
The oscillation periods presented in figure~\ref{fig.period} are obtained from a polynomial least square fit of the time-averaged alignment data.

\textbf{Simulations.}
We model the non-resonant laser-molecule interaction by the effective potential $U=-(\Delta\alpha/4)\cos^2\theta E^2(t)$, where $\Delta\alpha$ is the molecular polarisability anisotropy, and $E(t)$ is the envelope of the laser electric field, and $\theta$ is the angle between the molecular axis and the laser polarisation direction.
The wave function is expanded in the spherical harmonics, and we solve numerically the time-dependent Schr\"odinger equation to obtain the expansion coefficients.
To take into account thermal effects, we average over the initial states, where each result is weighted by the Boltzmann factor (including nuclear spin statistics) of the initial state.
A detailed description of the numerical method can be found in Ref.~\cite{fleischer12}.

\section*{Acknowledgements}

I.A. and J.F. appreciate many fruitful discussions related to the problem with Yaron Silberberg.
P.H.B. and A.K. received support from the AMOS program, Chemical Sciences, Geosciences, and Bioscienses Division, Basic Energy Sciences, Office of Science, U.S. Department of Energy.
I.A. and J.F. acknowledge financial support by the ISF (Grant No. 601/10), the DFG (Project No. LE 2138/2-1), and the Minerva Foundation.
I.A. acknowledges support as the Patricia Elman Bildner Professorial Chair.
This research was made possible in part by the historic generosity of the Harold Perlman Family.

\section*{Author contributions}

A.K. performed the experiments.
J.F. performed the numerical simulations.
I.A. and P.H.B. supervised the project.
All authors discussed the results and took part in preparing the manuscript.



\end{document}